\documentclass[12pt]{article}
\usepackage{epsfig}

\newcommand{\msun}{M_{\odot}}
\newcommand{\pkgr}{{P_{\,\rm K}}}
\newcommand{\okgr}{{\Omega_{\,\rm K}}}

\def\beq{\begin{equation}}
\def\eeq#1{\label{#1}\end{equation}}
\def\eeqn{\end{equation}}

\def\beqa{\begin{eqnarray}}
\def\eeqa#1{\label{#1}\end{eqnarray}}
\def\eeqan{\end{eqnarray}}

\let\bar=\overbar

\def\Dslash{\not{\hbox{\kern-4pt $D$}}}
\def\dslash{\not{\hbox{\kern-2pt $\del$}}}

\def\msb{{\bar{\ssstyle M \kern -1pt S}}}

\usepackage{fancyhdr,graphicx}
\def\Title#1{\begin{center} {\Large {\bf #1} } \end{center}}

\begin{document}

\Title{Impact of Rotation on the Structure and Composition of Neutron Stars}

\bigskip\bigskip


\begin{raggedright}

{\it Fridolin Weber}\\ Department of Physics, San Diego State
University, San Diego, California 92182, USA \& Center for
Astrophysics and Space Sciences, University of California, San Diego,
La Jolla, California 92093, USA \\ {\tt Email: fweber@mail.sdsu.edu}\\
{\it Milva Orsaria}\footnote{Home address: CONICET, Rivadavia 1917, 1033
Buenos Aires, Argentina;\\ Gravitation, Astrophysics and Cosmology Group, Facultad de Ciencias Astron{\'o}micas y Geof{\'i}sicas,\\ Paseo del Bosque S/N (1900), Universidad Nacional de La Plata UNLP, La Plata, Argentina} \\ Department of Physics, San Diego State
University, San Diego, CA 92182, USA \\ {\tt Email: morsaria@rohan.sdsu.edu}\\
{\it Rodrigo Negreiros} \\ Instituto de F\'isica, Universidade Federal
Fluminense, Niter\'oi, RJ, Brazil\\ {\tt Email: negreiros@if.uff.br}\\

\bigskip\bigskip
\end{raggedright}

\section{Introduction}

Depending on mass and rotational frequency, gravity compresses the
matter in the core regions of neutron stars to densities that are
several times higher than the density of ordinary atomic nuclei
\cite{glen97:book,weber99:book,blaschke01:trento,lattimer01:a,weber05:a,%
  page06:review,klahn06:a_short,sedrakian07:a,klahn07:a}. At such huge
densities atoms themselves collapse, and atomic nuclei are squeezed so
tightly together that new particle states may appear and novel states
of matter, foremost quark matter, may be created.  This feature makes
neutron stars superb astrophysical laboratories for a wide range of
physical studies \cite{glen97:book,weber99:book,blaschke01:trento,%
  lattimer01:a,page06:review,sedrakian07:a,alford08:a}.  And with
observational data accumulating rapidly from both orbiting and ground
based observatories spanning the spectrum from X-rays to radio
wavelengths, there has never been a more exiting time than today to
study neutron stars. The Hubble Space Telescope and X-ray satellites
such as Chandra and XMM-Newton in particular have proven especially
valuable.  New astrophysical instruments such as the Five hundred
meter Aperture Spherical Telescope (FAST), the square kilometer Array
(skA), Fermi Gamma-ray Space Telescope (formerly GLAST), and possibly
the International X-ray Observatory (now Advanced Telescope for High
ENergy Astrophysics, ATHENA) promise the discovery of tens of
thousands of new non-rotating and rotating neutron stars. The latter
are referred to as pulsars.  This paper provides a short overview of
the impact of rotation on the structure and composition of neutron
stars. Observational properties, which may help to shed light on the
core composition of neutron stars--and, hence, the properties of
ultra-dense matter \cite{CBMbook11:a,NICA09:a}--are discussed.

\section{Modeling of Rotating Neutron Stars}

\begin{figure}[htb]
\begin{center}
\epsfig{file=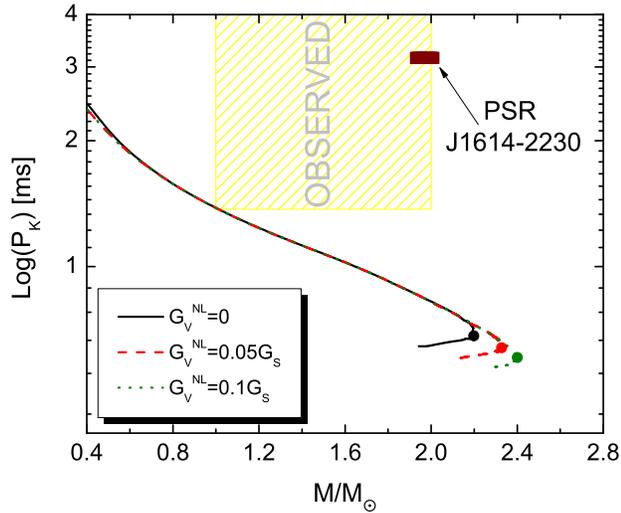,height=2.8in}
\caption{Kepler period, $\pkgr$, of a sequence of rotating neutron
  stars computed for the NJL equation of state discussed in Section
  \ref{sec:quark_matter}. The colored solid dots denote the
  termination point of each stellar sequence, which depends on the
  stiffness (i.e.\ $G_V^{NL}$ value) of the equation of state.}
\label{fig:Kepler}
\end{center}
\end{figure}

The structure equations of rotating neutron stars are based on a line
element of the form
\cite{weber99:book,friedman86:a,lattimer90:a,eriguchi93:a,salgado94:a,cook94:a,%
  cook94:b}
\begin{eqnarray}
  d s^2 = - e^{2\nu} dt^2 + e^{2\psi} (d\phi - \omega dt)^2 + e^{2\mu}
  d\theta^2 + e^{2\lambda} dr^2 \, ,
\label{eq:metric} 
\end{eqnarray} where each metric function,  $\nu$, $\psi$,  $\mu$ and
$\lambda$, as well as the angular velocities of the local inertial
frames, $\omega$, depend on the radial coordinate $r$ and on the polar
angle $\theta$ and, implicitly, on the star's angular velocity
$\Omega$.  The metric functions and the frame dragging frequency are
to be computed from Einstein's field equation, 
\begin{equation} 
G^{\alpha \beta} \equiv R^{\alpha\beta} - \frac{1}{2} \, R \,
g^{\alpha\beta} = 8 \, \pi \, T^{\alpha \beta} \, ,
\end{equation}
where $T^{\alpha \beta} = T^{\alpha
  \beta}(\epsilon, P(\epsilon))$ denotes the energy momentum tensor of
the stellar matter whose equation of state is given by $P(\epsilon)$.
No simple stability criteria are known for rapidly rotating stellar
configurations in general relativity. However, an absolute limit on
rapid rotation is set by the onset of mass shedding from the equator
of a rotating star. The corresponding rotational frequency is known as
the Kepler frequency, $\okgr$. In classical mechanics, the expression
for the Kepler frequency, determined by the equality between the
centrifugal force and gravity, is readily obtained as $\okgr =
\sqrt{M/R^3}$. Its general relativistic counterpart is given by
\cite{weber99:book,friedman86:a}
\begin{eqnarray}
  \okgr = \omega +\frac{\omega_{,r}} {2\psi_{,r}} + e^{\nu -\psi} \sqrt{
    \frac{\nu_{,r}} {\psi_{,r}} + \Bigl(\frac{\omega_{,r}}{2
      \psi_{,r}} e^{\psi-\nu}\Bigr)^2 } \, , 
\label{eq:okgr}  
\end{eqnarray} 
which is to be evaluated self-consistently at the equator of a
rotating neutron star. The Kepler period follows from
Eq.\ (\ref{eq:okgr}) as $\pkgr = {{2 \pi} / \okgr}$. For typical
neutron star matter equations of state, the Kepler period obtained for
$1.4\, \msun$ neutron stars is around 1~ms (Fig.\ \ref{fig:Kepler})
\cite{weber99:book,weber05:a,friedman86:a,lattimer90:a}. An exception to this are
strange quark matter stars. Since they are self-bound, they tend to
possess smaller radii than conventional neutron stars, which are bound
by gravity only. Because of their smaller radii, strange stars can
withstand mass shedding from the equator down to periods of around
0.5~ms \cite{glen92:crust,glen92:limit}.

A mass increase of up to $\sim 20$\% is typical for rotation at
$\okgr$. Because of rotation, the equatorial radii increase by several
kilometers, while the polar radii become smaller by several
kilometers. The ratio between both radii is around 2/3, except for
rotation close to the Kepler frequency. The most rapidly rotating,
currently known neutron star is pulsar PSR J1748-2446ad, which rotates
at a period of 1.39~ms (719~Hz) \cite{hessels06:a}, well below the
Kepler frequency for most neutron star equations of state. Examples of
other rapidly rotating neutron stars are PSRs B1937+21
\cite{backer82:a} and B1957+20 \cite{fruchter88:a}, whose rotational
periods are 1.58 ms (633~Hz) and 1.61~ms (621~Hz), respectively.

The density change in the core of a neutron star whose frequency
varies from $0 \leq \Omega \leq \okgr$ can be as large as 60\%
\cite{weber99:book,weber05:a}. This suggests that rotation may drive
phase transitions or cause significant compositional changes of the
matter in the cores of neutron stars.  Examples of the latter are
shown in Figs.\ \ref{fig:bonn_eq_po_1.40} and
\ref{fig:bonn_eq_po_1.70}, which illustrate rotation-driven changes in
the hyperon compositions of neutron stars.  Qualitatively similar
restructuring effects are obtained if neutron stars should contain
deconfined quark matter in their cores
\cite{glen97:book,weber99:book,weber05:a}. Quantitatively, however,
there may be a striking difference if quark matter is present, which
could lead to a ``backbending'' of pulsars \cite{glen97:a}. In the
latter case, quark deconfinement may register itself in the
braking index $n$, which could be somewhere between $- \infty < n < +
\infty$ rather than $n\approx 3$, as well as in the spin-ups of
isolated pulsars, which could last from tens of
thousands to hundreds of thousands of years \cite{weber05:a,glen97:a}.

\section{Core Composition of Rotating Neutron Stars}
\label{sec:core_composition}

The hadronic phase inside of neutron stars may be described in the
framework of non-linear relativistic nuclear field theory
\cite{glen97:book,weber99:book}, where baryons (neutrons, protons,
hyperons) interact via the exchange of scalar, vector and isovector
mesons ($\sigma$, $\omega $, $\vec \rho $, respectively).  The
Lagrangian of the theory is given by
\begin{eqnarray}
  \mathcal{L} &=& \sum_{B=n,p, \Lambda, \Sigma, \Xi}\bar{\psi}_B
  \bigl[\gamma_\mu(i\partial^\mu-g_\omega
  \omega^\mu-g_\rho \vec{\, \rho}^\mu) 
  -(m_N-g_\sigma\sigma)\bigr]\psi_B+\frac{1}{2}(\partial_\mu\sigma\partial^\mu
  \sigma-m_\sigma^2\sigma^2) \nonumber\\
  &-&\frac{1}{3}b_\sigma m_N(g_\sigma\sigma)^3 - \frac{1}{4}c_\sigma(g_\sigma\sigma)^4-
  \frac{1}{4}\omega_{\mu\nu}\omega^{\mu\nu} 
  +\frac{1}{2} m_\omega^2 \, \omega_\mu\omega^\mu + \frac{1}{2} m_\rho^2\,
  \vec{\rho}_\mu\cdot  \vec{\rho\,}^\mu \label{eq:lag}\\
  &-&\frac{1}{4} \vec{\rho}_{\mu\nu} \vec{\rho\,}^{\mu\nu} + \sum_{\lambda=e^-, \mu^-}
  \bar{\psi}_\lambda  (i\gamma_\mu\partial^\mu-m_\lambda)\psi_\lambda \, ,
  \nonumber
\end{eqnarray}
where $B$ sums all baryon states which become populated in neutron
star matter \cite{glen97:book,weber99:book}. The quantities
$g_\rho$, $g_\sigma$, and $g_\omega$ are the meson-baryon coupling
constants. Non-linear $\sigma$-meson self-interactions are taken into
account in Eq.\ (\ref{eq:lag}) via the terms proportional to
$b_\sigma$ and $c_\sigma$. The
equations of motion for the baryon and meson field equations, which
follow from Eq.\ (\ref{eq:lag}), can be solved using the relativistic
mean-field approximation, where
\begin{figure}[tb]
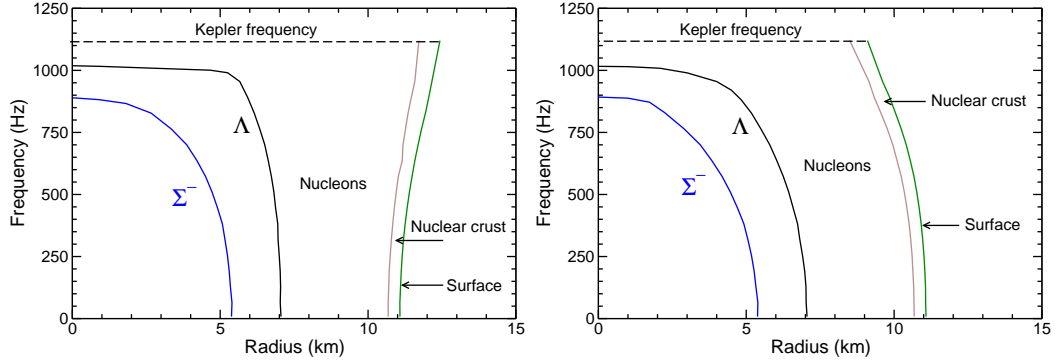

\begin{center}
\begin{tabular}{cc}
\includegraphics[width=0.45\textwidth]{1.40_eq_bonn.eps} 
\includegraphics[width=0.45\textwidth]{1.40_po_bonn.eps}
\end{tabular} 
\caption{Composition of a rotating neutron star in equatorial
  direction (left panel) and polar direction (right panel)
  \cite{weber99:book,weber07:HYP}. The star's mass at zero rotation is
  $1.40\, \msun$.}
\label{fig:bonn_eq_po_1.40}
\end{center}
\end{figure}
the meson fields are approximated by their respective mean-field
values $\bar{\sigma} \equiv \langle\sigma\rangle$, $\bar\omega \equiv
\langle\omega\rangle$, and ${\bar\rho_{03}} \equiv
\langle\rho_{03}\rangle$ \cite{glen97:book,weber99:book}.  The
\begin{figure}[tb]
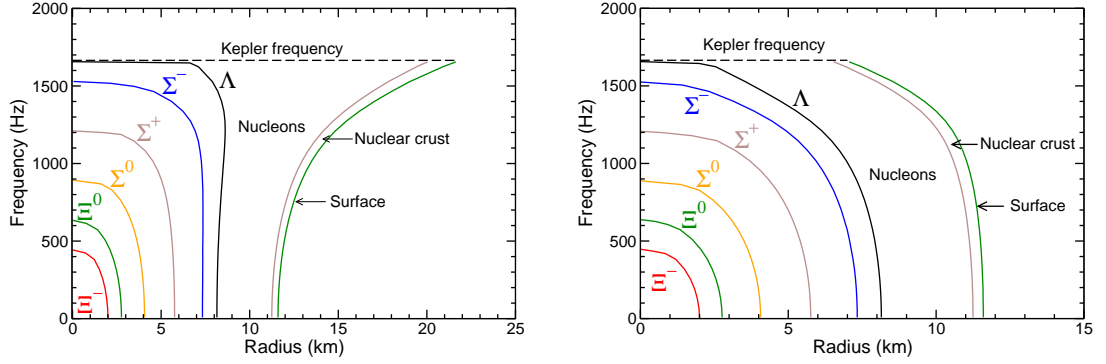

\begin{center}
\begin{tabular}{cc}
\includegraphics[width=0.45\textwidth]{1.70_eq_bonn.eps} ~~&
\includegraphics[width=0.45\textwidth]{1.70_po_bonn.eps}
\end{tabular} 
\caption{Same as Figure \ref{fig:bonn_eq_po_1.40}, but for a
  non-rotating stellar mass of $1.70\, \msun$
  \cite{weber99:book,weber07:a}.}
\label{fig:bonn_eq_po_1.70}
\end{center}
\end{figure}
parameters of the model are adjusted to the properties of nuclear
matter at saturation density.  Figures\ \ref{fig:bonn_eq_po_1.40} and
\ref{fig:bonn_eq_po_1.70} show the compositions of rotating neutron
stars based on the Lagrangian of Eq.\ (\ref{eq:lag}). One sees that,
depending on the mass of a neutron, certain hyperon species may not be
present at high neutron star rotation rates.

\section{Quark Matter in the Cores of Neutron Stars}
\label{sec:quark_matter}

Whether or not quark matter exists in the cores of neutron stars is an
open issue. The observation of neutron stars with masses of up to
around $2\, \msun$ does not yet rule out quark matter
\cite{alford06:b,orsaria13:a}. To model quark deconfinement, one may
start from the Euclidean effective action associated with the nonlocal
SU(3) quark model (for details, see \cite{orsaria13:a} and references
therein),
\begin{eqnarray}
S_E &=& \int d^4x \ \{ \bar \psi (x) \left[ -i \gamma_\mu
\partial_\mu + \hat m \right] \psi(x)
 - \frac{G_s}{2} \left[
j_a^S(x) \ j_a^S(x) + j_a^P(x) \ j_a^P(x) \right]\nonumber \\
&-& \frac{H}{4} \ T_{abc} \left[
j_a^S(x) j_b^S(x) j_c^S(x) - 3\ j_a^S(x) j_b^P(x) j_c^P(x) \right]
 - \frac{G_{V}^{NL}}{2} j_{V,f}^\mu(x) j_{V,f}^\mu(x) \, , \label{se}
\end{eqnarray}
where $\psi$ is a chiral $U(3)$ vector that includes the light quark
fields, $\psi \equiv (u\; d\; s)^T$, and $\hat m = {\rm diag}(m_u,
m_d, m_s)$ stands for the current quark mass matrix. For simplicity we
consider the isospin symmetry limit so that  $m_u = m_d=\bar m$. The
fermion kinetic term includes the covariant derivative $D_\mu\equiv
\partial_\mu - iA_\mu$, where $A_\mu$ are color gauge fields, and the
operator $\gamma_\mu\partial_\mu$ in Euclidean space is defined as
$\vec \gamma \cdot \vec \nabla + \gamma_4\frac{\partial}{\partial
  \tau}$, with $\gamma_4=i\gamma_0$. The currents $j_a^{S,P}(x)$ and
$j_{V,f}^{\mu}(x)$ are defined in \cite{orsaria13:a}.  Using the
mean-field approximation, one obtains from Eq.\ (\ref{se}) the grand
canonical potential
\begin{eqnarray}
  && \Omega^{NL} ( T=0,\mu_f) =  -\,\frac{N_c}{\pi^3}\,\sum_{f=u,d,s}\, \int^{\infty}_{0}
  \,dp_0 \int^{\infty}_{0}\,dp 
   \mbox{ ln }\left\{\left[\omega_f^2 + M_{f}^2(\omega_f^2)\right]\,\frac{1}{\omega_f^2
      + m_{f}^2}\right\} \nonumber \\
  && -\,\frac{N_c}{\pi^2}\, \sum_{f=u,d,s}\,\int^{\sqrt{ \widetilde{\mu}_f^2-m_{f}^2}}_{0} \,
  dp\,p^2 \,
  \left[\,(\widetilde{\mu}_f-E_f)\, \theta(\widetilde{\mu}_f-m_f)\,\right]\nonumber \\
  &&  -\; \frac{1}{2}\left[ \sum_{f=u,d,s} (\bar \sigma_f \ \bar S_f  +
    \frac{G_s}{2} \ \bar S_f^2) \; + \; \frac{H}{2} \, \bar S_u\ \bar S_d\ \bar S_s
  \right] 
  -\;\sum_{f=u,d,s} \frac{{ \varpi^2_{V,f}}}{4 G_V^{NL}}, \label{omzerot}
\end{eqnarray}
where $N_c=3$, $E_{f}=\sqrt{\vec{p}\,^{2}+m_{f}^{2}}$, and $\omega_f^2
= (\,p_0\, + \,i\, \mu_f\,)^2\, + \,\vec{p}\,^2$.  The masses of free
quarks are denoted by $m_f$, where $f=u,d,s$.  The momentum-dependent
constituent quark masses $M_{f}$ depend explicitly on the quark mean
fields $\bar\sigma_{f}$, $ M_{f}(\omega_{f}^2) \ = \ m_f\, + \,
\bar\sigma_f\, g(\omega_{f}^2)$, where $g(\omega^2)$ denotes the
Fourier transform of the form factor $\widetilde{g}(z)$. For a
Gaussian form factor one has $g(\omega^2) =
\exp{\left(-\omega^2/\Lambda^2\right)} $, where $\Lambda$ plays a
role for the stiffness of the chiral transition. This parameter,
together with the current quark mass
\begin{figure}[tb]
\begin{center}
\begin{tabular}{cc}
\includegraphics[width=0.51\textwidth]{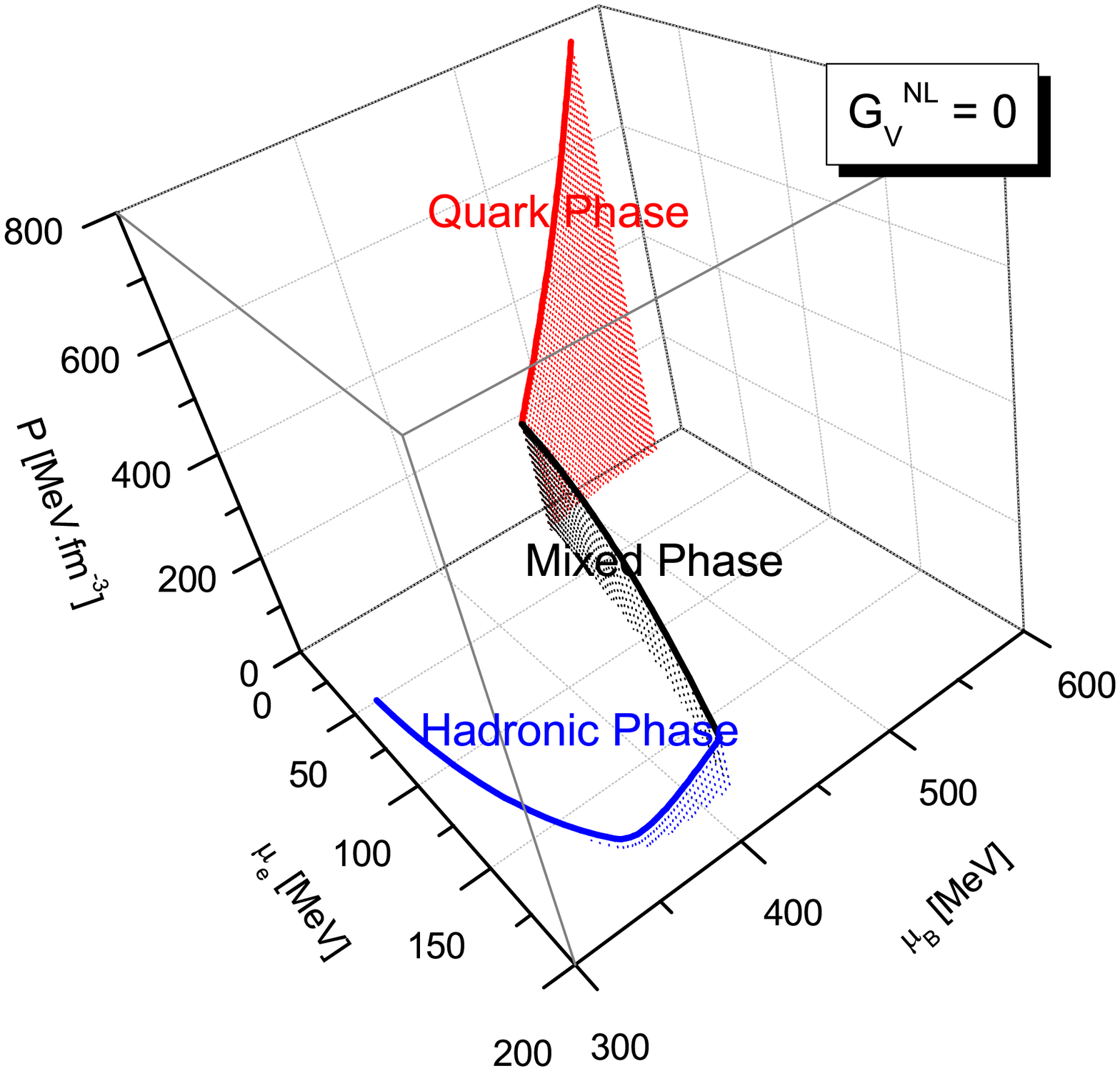} &
\includegraphics[width=0.51\textwidth]{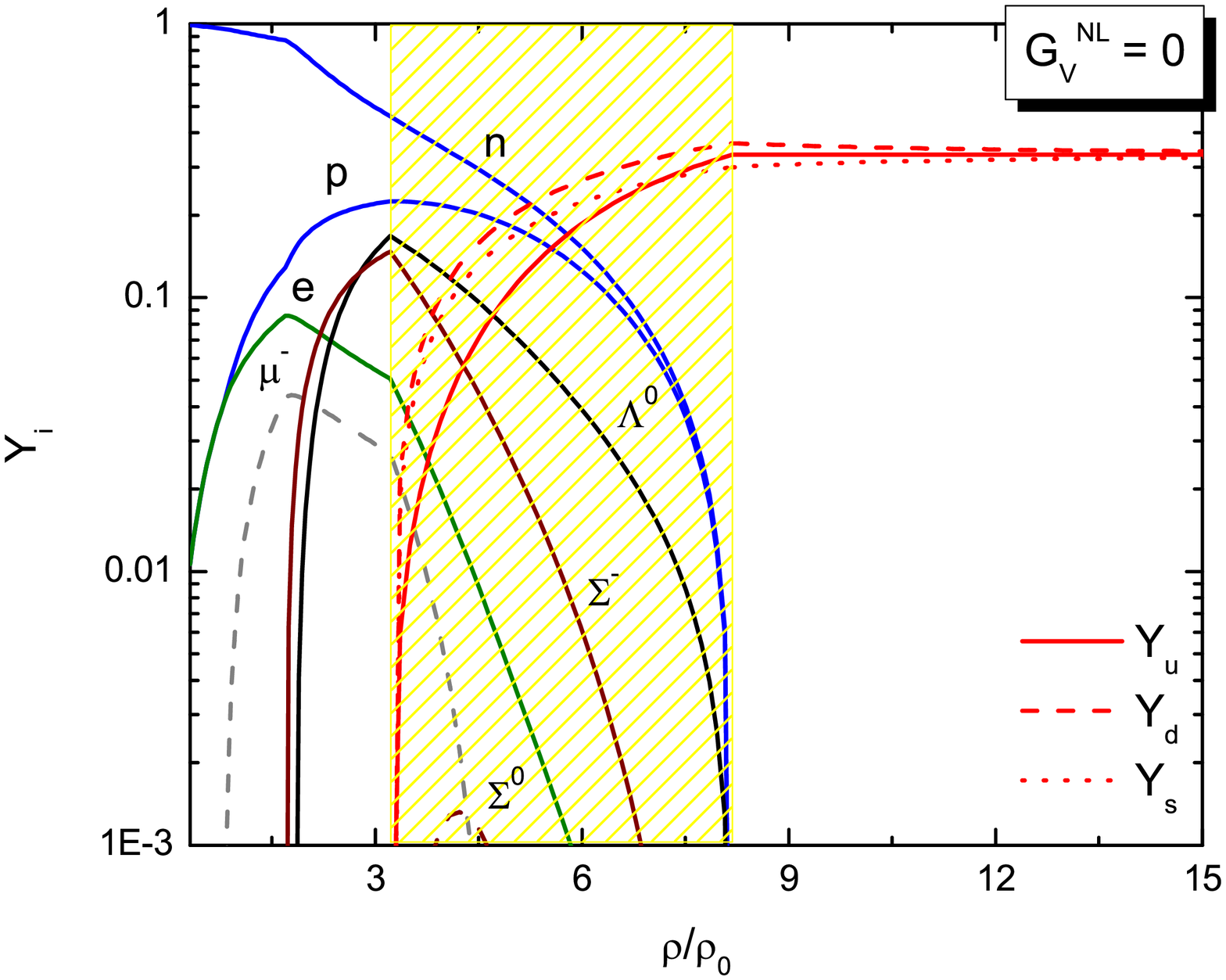}
\end{tabular} 
\caption{Equation of state (left panel) and associated quark-lepton
  composition (right panel) computed for $G_V^{NL}=0$.}
\label{fig:GV0}
\end{center}
\end{figure}
$\bar{m}$ of up and down quarks and the coupling constants $G_s$ and
$H$ in Eq.\ (\ref{omzerot}), have been fitted to the pion decay
constant, $f_\pi$, and meson masses $m_{\pi}$, $m_\eta$, and
$m_{\eta'}$ \cite{orsaria13:a}. The result of this fit is $\bar{m} =
6.2$~MeV, $\Lambda = 706.0$~MeV, $G_s \Lambda^2 = 15.04$, $H \Lambda^5
= - 337.71$. The strange quark current mass is treated as a free
parameter and was set to $m_s =140.7$~MeV. Using these parametrizations, the fields
$\bar \sigma_f$ and $\varpi_{V,f}$ can be determined by minimizing
Eq.\ (\ref{omzerot}), $ \partial \Omega^{NL} / \partial \bar \sigma_f
= \partial \Omega^{NL} / \partial {\varpi_{V,f}}= 0$. The strength of the vector
interaction $G_V$ is usually expressed in terms of the strong coupling
constant $G_s$. To account for the uncertainty in the theoretical
predictions for the ratio $G_V/G_s$, the vector coupling constant may
be treated as a free parameter which varies
from $0$ to $0.1 \, G_s$.  

To model the mixed phase region of quarks and hadrons in neutron
stars, we use the Gibbs condition for phase equilibrium between
hadronic ($H$) and quark ($Q$) matter,
\begin{eqnarray}
  P_H ( \mu_n , \mu_e, \{ \phi \} ) = P_Q (\mu_n , \mu_e) \, ,
\label{eq:gibbs}
\end{eqnarray}
where $P_H$ and $P_Q$ denote the pressures of hadronic matter and
quark matter, respectively \cite{glen91:pt,glen01:b}. The
quantity $\{ \phi \}$ in Eq.\ (\ref{eq:gibbs}) stands collectively for
the field variables ($\bar{\sigma}$, $\bar\omega$, $\bar\rho$) and
Fermi momenta ($k_B$, $k_\lambda$) that characterize a solution to the
equations of confined hadronic matter
(Sect.\ \ref{sec:core_composition}). We use the symbol $\chi \equiv
V_Q/V$ to denote the volume proportion of quark matter, $V_Q$, in the
unknown volume $V$. By definition, $\chi$ then varies between 0 and 1,
depending on how much confined hadronic matter has been converted to
quark matter.  Equation (\ref{eq:gibbs}) is to be supplemented with
the conditions of global baryon charge conservation and global
electric charge conservation.  The global conservation of baryon
charge is expressed as \cite{glen91:pt,glen01:b}
\begin{eqnarray}
  \rho_b = \chi \, \rho_Q(\mu_n, \mu_e ) + (1-\chi) \,
  \rho_H (\mu_n, \mu_e,  \{ \phi \}) \, ,
\label{eq:mixed_rho}
\end{eqnarray}
where $\rho_Q$ and $\rho_H$ denote the baryon number densities of the
quark phase and hadronic phase, respectively. The global neutrality of
electric charge is given by \cite{glen91:pt,glen01:b}
\begin{eqnarray}
  0 = \chi \ q_Q(\mu_n, \mu_e ) + (1-\chi) \ q_H (\mu_n, \mu_e, \{
  \phi \}) \, ,
\label{eq:mixed_charge}
\end{eqnarray}
with $q_Q$ and $q_H$ denoting the electric charge densities of the
quark phase and hadronic phase, respectively.  We have chosen global
rather than local electric charge neutrality.  Local NJL studies
carried out for local electric charge neutrality have been reported
recently in Refs.\ \cite{masuda12:a,lenzi12:a}. Figures \ref{fig:GV0}
shows a model quark hadron composition and its associated equation of
state computed for the nonlocal NJL model.
\begin{figure}[tb]
\begin{center}
\epsfig{file=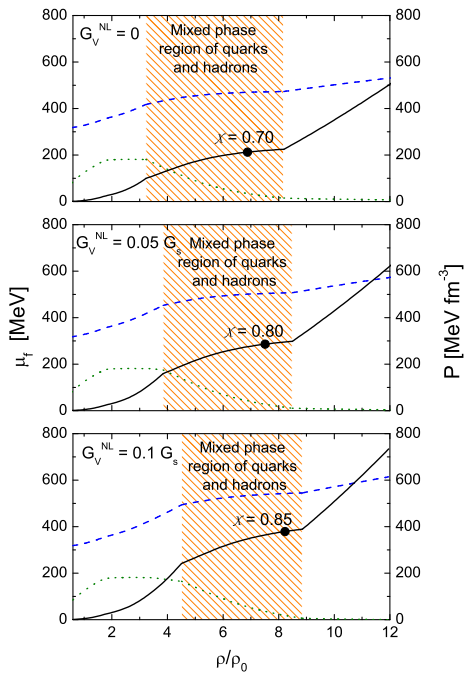,height=3.6in}
\caption{Pressure, $P$ (solid lines), baryon chemical potential,
  $\mu_b$ (dashed lines), and electron chemical potential, $\mu_e$
  (dotted lines) as a function of baryon number density, $\rho$, in
  units of the normal nuclear matter density, $\rho_0 = 0.16~{\rm
    fm}^{-3}$. The hatched areas denote the mixed phase regions where
  confined hadronic matter and deconfined quark matter coexist. The
  solid dots indicate the central densities of the associated
  maximum-mass stars, and $\chi$ is the respective fraction of quark
  matter inside of them. The results are shown for three different
  values of the vector coupling constant, ranging from 0, to 0.05
  $G_s$, to 0.1 $G_s$. (Figure from Ref.\ \cite{orsaria13:a}.)}
\label{press}
\end{center}
\end{figure}
A model composition for the mixed phase region is shown in Fig.\ \ref{press}. The inclusion of
the quark vector coupling contribution shifts the onset of the phase
transition to higher densities, and also narrows the width of the
mixed quark-hadron phase, when compared to the case $G_V = 0$. The
mixed phases range from $3.2 - 8.2 \rho_0$, $3.8 - 8.5 \rho_0$, and
$4.5 - 8.9 \rho_0$ for vector coupling constants $G_V/G_s = 0, \ 0.05
, \ 0.1$, respectively. We note that there is considerable theoretical
uncertainty in the ratio of $G_{V}/G_{s}$ \cite{dumm07:a} since a
rigorous derivation of the effective couplings from QCD is not
possible. Combining the ratios of $G_{V}/G_{s}$ from the molecular
instanton liquid model and from the Fierz transformation, the value of
$G_{V}/G_{s}$ is expected to be in the range $0 \leq G_{V}/G_{s} \leq
0.5$ \cite{zhang09:a}. For our model, values of $G_{V}/G_{s} > 0.1$
shift the onset of the quark-hadron phase transition to such high
densities that quark deconfinement can not longer occur in the cores
of neutron stars.

As shown in \cite{orsaria13:a}, the maximum neutron star masses increase from
$1.87 \, M_\odot$ for $G_V = 0$, to $2.00 \, M_\odot$ for $G_V = 0.05
\, G_s$, to $2.07 \,M_\odot$ for $G_V = 0.1 \, G_s$. The heavier stars
of all three stellar sequences contain mixed phases of quarks and
hadrons in their centers. The densities in these stars are however not
high enough to generate pure quark matter in the cores. Such matter
forms only in neutron stars which are already located on the
gravitationally unstable branch of the mass-radius
relationships. Another intriguing feature of this model is that
neutron stars with canonical masses of around 1.4 $M_\odot$ do not
posses a mixed phase of quarks and hadrons but are made entirely of
confined hadronic matter.

\section{Thermal evolution of rotating neutron stars}
\label{ssec:nutshell}

The basic cooling features of a neutron star are easily grasped by
considering the energy conservation relation of the star in the
Newtonian limit \cite{page05:a}. This equation is given by
\begin{eqnarray}
dE_{\rm th}/dt= C_V dT/dt = -L_\nu - L_\gamma + H \, , 
\label{eq:cooling1}
\end{eqnarray}
where $E_{\rm th}$ is the thermal energy content of a neutron star,
$T$ its internal temperature, and $C_V$ its total specific heat. The
energy sinks are the total neutrino luminosity, $L_\nu$, and the
surface photon luminosity, $L_\gamma$. The source term $H$ includes
all possible heating mechanisms \cite{page05:a}, which, for instance,
convert magnetic or rotational energy into heat. The dominant
contributions to $C_V$ come from the core whose constituents are
leptons, baryons, boson condensates and possibly deconfined
superconducting quarks.  When baryons and quarks become paired, their
contribution to $C_V$ is strongly suppressed at temperatures smaller
than the critical temperatures associated with these pairing
phases. The crustal contribution is in principle dominated by the free
neutrons in the inner stellar crust but, since these are extensively
paired, practically only the nuclear lattice and electrons
contribute. Extensive baryon, and quark, pairing can thus
significantly reduce $C_V$. In order to derive the general
relativistic version of Eq.\ (\ref{eq:cooling1}) for rotating stars,
one needs to solve Einstein's field equations using the metric of a
rotationally deformed fluid defined in Eq.\ (\ref{eq:metric}). The
result is the following parabolic differential equation,
\begin{eqnarray}
  \partial_t \tilde T &=& - \, {1 \over{\Gamma^2}} e^{2 \nu} {\epsilon \over{
      C_V}} - r \sin \theta \, U e^{\nu+\gamma-\xi} {1 \over{C_V}}
\left( \partial_r \Omega + {1 \over r} \partial_\theta \Omega \right) 
\nonumber \\
&& + {1 \over{r^2 \sin\theta}} {1 \over \Gamma} e^{3\nu-\gamma-2\xi}
{1 \over {C_V}} \Bigl( \partial_r \left( r^2 \kappa \sin \theta \, e^\gamma
\left( \partial_r \tilde T + \Gamma^2 U e^{-2\nu+\gamma} \tilde T \partial_r
  \Omega 
\right) \right)  \nonumber \\
&& + {1 \over{r^2}} \partial_\theta \left( r^2 \kappa \sin \theta \, e^\gamma
\left( \partial_\theta \tilde T + \Gamma^2 U e^{-2\nu+\gamma} \tilde
  T \partial_\theta \Omega \right) \right) \Bigr) \, ,
\label{eq:3.49}
\end{eqnarray}
with the definitions $r \sin\theta e^{-\nu+\gamma} = e^\phi$,
$e^{-\nu+\xi} = e^{\alpha-\beta}$ and the Lorentz factor $\Gamma
\equiv (1 - U^2)^{-1/2}$.  This differential equation needs be solved
numerically in combination with a 2-dimensional general relativistic
stellar rotation code.  The latter is used to determine the metric
functions, frame dragging frequency, pressure and density gradients,
and particle composition of a deformed neutron star as a function of
its rotational frequency, $0 < \Omega \leq \okgr$. Depending on the
cooling channels that are active at a given frequency, the numerical
outcome of the rotation code serves as an input for the thermal
evolution code, which is used to determine the luminosity, and thus
the surface temperatures, of a deformed rotating neutron star. As
mentioned in Sect.\ \ref{sec:core_composition}, because of stellar
spin-down or spin-up, the density in the cores of rotating neutron
stars may change dramatically so that new cooling channels open up (or
close) with time. The changing cooling channels render the neutrino
emissivities, heat capacities, and thermal conductivities rotation
dependent, which alters the thermal response of the star, It is this
response in the thermal behavior of rotating neutron stars that
carries information about the properties of the matter in the dense
baryonic stellar cores and possibly the deep crustal layers.

In passing we mention that the general relativistic equations of
energy balance and thermal energy transport of non-rotating stars are
given by
\begin{eqnarray}
  \frac{ \partial (l e^{2\phi})}{\partial m}& = 
  &-\frac{1}{\rho \sqrt{1 - 2m/r}} \left( \epsilon_\nu 
    e^{2\phi} + c_v \frac{\partial (T e^\phi) }{\partial t} \right) \, , 
  \label{coeq1}  \\
  \frac{\partial (T e^\phi)}{\partial m} &=& - 
  \frac{(l e^{\phi})}{16 \pi^2 r^4 \kappa \rho \sqrt{1 - 2m/r}} 
  \label{coeq2} 
  \, ,
\end{eqnarray}
respectively \cite{weber99:book}. Here, $r$ is the distance from the
center of the star, $m(r)$ is the mass, $\rho(r)$ is the energy
density, $T(r,t)$ is the temperature, $l(r,t)$ is the luminosity,
$\phi(r)$ is the gravitational potential, $\epsilon_\nu(r,T)$ is the
neutrino emissivity, $c_v(r,T)$ is the specific heat, and
\begin{figure}[htb]
\begin{center}
 \includegraphics[width=0.65\textwidth]{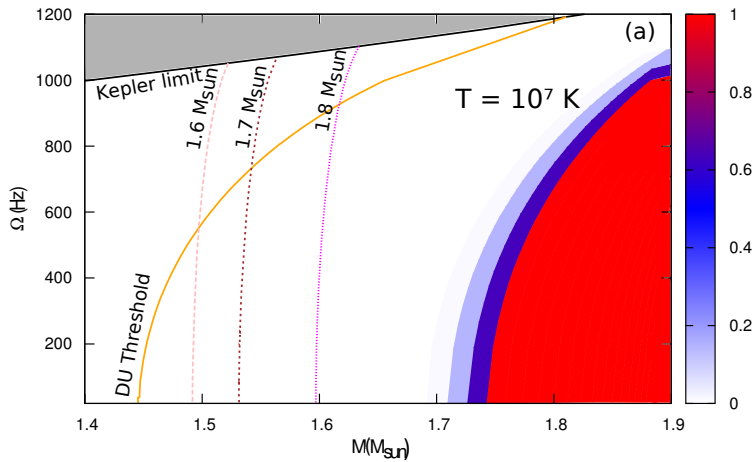}
\caption{Direct Urca (DU) process in rotating neutron stars of
  gravitational mass $M$ \cite{negreiros13:a}.  No stars are allowed
  above the curve labeled Kepler limit because of stellar mass
  shedding.  The vertical curves show evolutionary tracks of isolated
  rotating neutron stars, whose baryon mass remain constant during
  spin-down. One sees that for several stars evolving along these
  tracks the DU process (see text) is not allowed at high rotation
  rates but becomes operative at lower rotation rates.} 
  \label{Omgxfreq}
\end{center}
\end{figure}
$\kappa(r,T)$ is the thermal conductivity. The boundary conditions of
Eqs.\ (\ref{coeq1}) and (\ref{coeq2}) are determined by the luminosity
at the stellar center and at the surface. The luminosity vanishes at
the stellar center since there is no heat flux there.  At the surface,
the luminosity is defined by the relationship between the mantle
temperature and the temperature outside of the star.  Equations
(\ref{coeq1}) and (\ref{coeq2}) have been solved numerically in
\cite{niebergal10:a} for hypothetical color-flavor-locked
strange quark matter stars. It was found that such stars may undergo a
significant reheating because of the magnetic recombination of
rotational vortices, expelled from such stars during slow stellar
spin-down.

The full 2-dimensional cooling equations of rotating neutron stars
were solved recently in Ref.\ \cite{negreiros12:a} for isolated
rotating neutron stars.  Driven by the loss of energy, such stars are
gradually slowing down to lower frequencies, which increases the
tremendous compression of the matter inside of them. This increase in
compression changes both the global properties of rotating neutron
stars as well as their hadronic core compositions. Both effects may
register themselves observationally in the thermal evolution of such
stars, as demonstrated in \cite{negreiros13:a}.  The rotation-driven
particle process considered there was the direct Urca (DU) process $n
\rightarrow p + e^- + \bar\nu_e$ which becomes operative in neutron
stars if the number of protons in the stellar core exceeds a critical
limit of around 11\% to 15\% \cite{lattimer91:a}.  The neutrino
luminosity associated with this reaction dominates over those of other
neutrino emitting processes in the core \cite{page05:a,lattimer91:a}.
It was found that neutron stars spinning down from moderately high
rotation rates of a few
\begin{figure}[tb]
\begin{center}
\epsfig{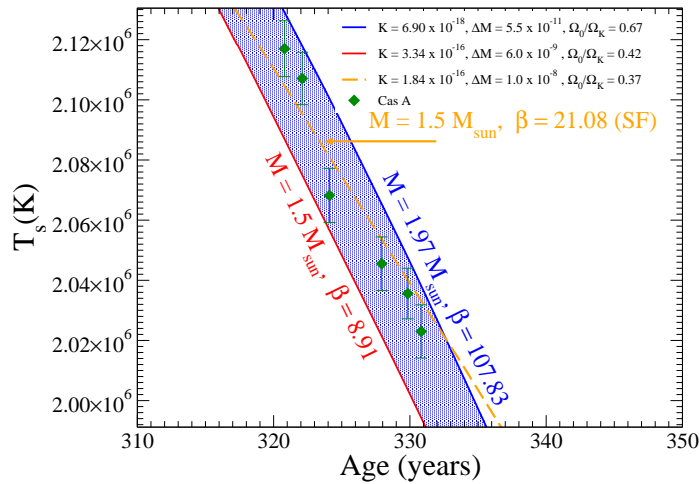}
\caption{Cooling simulations reproducing the temperatures observed for
  the neutron star in Cas~A over a ten-year period.  $\Delta M$
  denotes the mass of the accreted envelope, and $\Omega_0 / \Omega_K$
  is the star's birth frequency relative to the Kepler
  frequency. Superfluidity and Pair-Breaking formation of nucleons is
  taken into account in the curve labeled SF.}
\label{fig:CasA}
\end{center}
\end{figure}
hundred Hertz may be creating just the right conditions where the DU
process becomes operative (Fig.\ \ref{Omgxfreq}), leading to an
observable effect (enhanced cooling) in the temperature evolution of
such neutron stars. As it turned out (Fig.\ \ref{fig:CasA}), the
rotation-driven DU process could explain the unusual temperature
evolution observed for the neutron star in Cas~A, provided the mass of
this neutron star lies in the range of 1.5 to 1.9 $\msun$ and its
rotational frequency at birth was between 40 (400 Hz) and 70\% (800
Hz) of the Kepler (mass shedding) frequency, respectively
\cite{negreiros13:a}.

\bigskip

\noindent
{\bf Acknowledgment:} This material is based upon work supported by the
National Science Foundation under Grant No.\ 0854699.  M.\ Orsaria
thanks CONICET for financial support.


\begin{thebibliography}{10}

\bibitem{glen97:book}
N. K. Glendenning, {\it Compact Stars, Nuclear Physics, Particle Physics, and
  General Relativity}, 2nd ed.\ (Springer-Verlag, New York, 2000).

\bibitem{weber99:book}
F. Weber, {\it Pulsars as Astrophysical Laboratories for Nuclear and Particle
  Physics}, High Energy Physics, Cosmology and Gravitation Series (IOP
  Publishing, Bristol, Great Britain, 1999).

\bibitem{blaschke01:trento}
{\it Physics of Neutron Star Interiors}, ed.\ by D.\ Blaschke, N.\ K.\
  Glendenning, and A.\ Sedrakian, Lecture Notes in Physics {\bf 578}
  (Spring-Verlag, Berlin, 2001).

\bibitem{lattimer01:a}
J. M. Lattimer and M. Prakash, Astrophys.\ J.\ {\bf 550} (2001) 426.

\bibitem{weber05:a}
F. Weber, Prog.\ Part.\ Nucl.\ Phys.\ {\bf 54} (2005) 193.

\bibitem{page06:review}
D. Page and S. Reddy, Ann.\ Rev.\ Nucl.\ Part.\ Sci.\ {\bf 56} (2006) 327.

\bibitem{klahn06:a_short}
T. Kl{\"{a}}hn {\it et al.}, Phys.\ Rev.\ C {\bf 74} (2006) 035802.

\bibitem{sedrakian07:a}
A. Sedrakian, Prog.\ Part.\ Nucl.\ Phys.\ {\bf 58} (2007) 168.

\bibitem{klahn07:a}
T. Kl{\"{a}}hn {\it et al.}, Phys.\ Lett.\ B {\bf 654} (2007) 170.

\bibitem{alford08:a}
M. G. Alford, A. Schmitt, K. Rajagopal, and T. Sch{\"{a}}fer, Rev.\ Mod.\
  Phys.\ {\bf 80} (2008) 1455.

\bibitem{CBMbook11:a}
{\it Strongly Interacting Matter - The CBM Physics Book}, Lecture Notes in
  Physics {\bf 814}, 960 pages, (Springer, 2011).

\bibitem{NICA09:a}
NICA white paper, {\tt nica.jinr.ru/files/WhitePaper.pdf}.

\bibitem{friedman86:a}
J. L. Friedman, J. R. Ipser, and L. Parker, Astrophys.\ J.\ {\bf 304} (1986)
  115.

\bibitem{lattimer90:a}
J. M. Lattimer, M. Prakash, D. Masak, and A. Yahil, Astrophys.\ J.\ {\bf 355}
  (1990) 241.

\bibitem{eriguchi93:a}
Y. Eriguchi, in: {\it Rotating objects and relativistic physics}, ed.\ by F.\
  J.\ Chinea and L.\ M.\ Gonz{\'{a}}lez--Romero (Springer-Verlag, Berlin,
  1993).

\bibitem{salgado94:a}
M. Salgado, S. Bonazzola, E. Gourgoulhon, and P. Haensel, Astron.\ {\&}
  Astrophys.\ {\bf 291} (1994) 155.

\bibitem{cook94:a}
G. B. Cook, S. L. Shapiro, S. A. Teukolsky, Astrophys.\ J.\ {\bf 422} (1994)
  227.

\bibitem{cook94:b}
G. B. Cook, S. L. Shapiro, S. A. Teukolsky, Astrophys.\ J.\ {\bf 424} (1994)
  823.

\bibitem{glen92:crust}
N. K. Glendenning and F. Weber, Astrophys.\ J.\ {\bf 400} (1992) 647.

\bibitem{glen92:limit}
N. K. Glendenning, Phys.\ Rev.\ D {\bf 46} (1992) 4161.

\bibitem{hessels06:a}
J. W. T. Hessels, S. M. Ransom, I. H. Stairs, P. C. C. Freire, V. M. Kaspi, and
  F. Camilo, Science {\bf 311} (2006) 1901.

\bibitem{backer82:a}
D. C. Backer, S. R. Kulkarni, C. Heiles, M. M. Davis, and W. M. Goss, Nature
  {\bf 300} (1982) 615.

\bibitem{fruchter88:a}
A. S. Fruchter, D. R. Stinebring, and J. H. Taylor, Nature {\bf 334} (1988)
  237.

\bibitem{glen97:a}
N. K. Glendenning, S. Pei, and F. Weber, Phys.\ Rev.\ Lett.\ {\bf 79} (1997)
  1603.

\bibitem{weber07:HYP}
F. Weber and P. Rosenfield, {\it Rotating neutron stars}, Proceedings of HYP
  2006, ed. by J. Pochodzalla and Th. Walcher (Springer, Berlin, 2007) p.\ 381.

\bibitem{weber07:a}
F. Weber, R. Negreiros, P. Rosenfield, and M. Stejner, Prog.\ Part.\ Nucl.\
  Phys.\ 59 (2007) 94.

\bibitem{alford06:b}
M. Alford, D. Blaschke, A. Drago, T. Kl{\"{a}}hn, G. Pagliara, J.
  Schaffner-Bielich, {\it Quark matter and the masses and radii of compact
  stars}, ({\tt astro-ph/0606524}).

\bibitem{orsaria13:a}
M. Orsaria, H. Rodrigues, F. Weber, and G. A. Contrera, Phys.\ Rev.\ D {\bf 87}
  (2013) 023001.

\bibitem{glen91:pt}
N. K. Glendenning, Phys.\ Rev.\ D {\bf 46} (1992) 1274.

\bibitem{glen01:b}
N. K. Glendenning, Phys.\ Rep.\ {\bf 342} (2001) 393.

\bibitem{masuda12:a}
K. Masuda, T. Hatsuda and T. Takatsuka, {\tt arXiv:1205.3621 [nucl-th].}

\bibitem{lenzi12:a}
C. H. Lenzi and G. Lugones, Astrophys. J. {\bf 759} (2012) 57.

\bibitem{dumm07:a}
D. B. Blashcke, D. Gomez Dumm, A. G. Grunfeld, T. Kl\"{a}hn, and N. N.
  Scoccola, Phys. \ Rev.\ C {\bf 75} (2007) 065804.

\bibitem{zhang09:a}
Z. Zhang and T. Kunihiro, Phys.\ Rev. \ D {\bf 80} (2009) 014015.

\bibitem{page05:a}
D. Page, U. Geppert, and F. Weber, Nucl. Phys. {\bf A 777} (2006) 492.

\bibitem{negreiros13:a}
R. Negreiros, S. Schramm, and F. Weber, Phys.\ Lett.\ B {\bf 718} (2013) 1176.

\bibitem{niebergal10:a}
B. Niebergal, R. Ouyed, R. Negreiros, and F. Weber, Phys.\ Rev.\ D {\bf 81}
  (2010) 043005.

\bibitem{negreiros12:a}
R. Negreiros, S. Schramm, and F. Weber, Phys.\ Rev.\ D\ {\bf 85} (2012) 104019.

\bibitem{lattimer91:a}
J. M. Lattimer, C. J. Pethick, M. Prakash, and P. Haensel, Phys.\ Rev.\ Lett.\
  {\bf 66} (1991) 2701.

\end{thebibliography}

\end{document}